\documentclass[sigconf,authorversion,nonacm]{acmart}
\usepackage{subcaption}

\AtBeginDocument{%
  \providecommand\BibTeX{{%
    \normalfont B\kern-0.5em{\scshape i\kern-0.25em b}\kern-0.8em\TeX}}}

\makeatletter
\def\@copyrightspace{\relax}
\makeatother

\begin{document}

%%
%% The "title" command has an optional parameter,
%% allowing the author to define a "short title" to be used in page headers.

\title{Automatically Selecting Striking Images for Social Cards}

%%
%% The "author" command and its associated commands are used to define
%% the authors and their affiliations.
%% Of note is the shared affiliation of the first two authors, and the
%% "authornote" and "authornotemark" commands
%% used to denote shared contribution to the research.
\author{Shawn M. Jones}
% \affiliation{Los Alamos National Laboratory}
% \affiliation{Los Alamos, NM, USA}
% \affiliation{\city{Los Alamos}\state{New Mexico}\country{USA}}
% \affiliation{\url{https://orcid.org/0000-0002-4372-870X}}
\email{smjones@lanl.gov}
\orcid{0000−0002−4372−870X}
\affiliation{%
  \institution{Los Alamos National Laboratory}
  \streetaddress{P.O. Box 1663}
  \city{Los Alamos}
  \state{New Mexico}
  \country{USA}
  \postcode{87545}
  \\\url{https://orcid.org/0000-0002-4372-870X}
}

% \authornote{Both authors contributed equally to this research.}

\author{Michele C. Weigle}
% \affiliation{Old Dominion University}
% \affiliation{Norfolk, VA, USA}
% \affiliation{\city{Norfolk}\state{Virginia}\country{USA}}
% \affiliation{\url{https://orcid.org/0000-0002-2787-7166}}
\email{mweigle@cs.odu.edu}
\orcid{0000−0002−2787−7166}
% \authornotemark[1]
\affiliation{%
  \institution{Old Dominion University}
  \streetaddress{1 Old Dominion University}
  \city{Norfolk}
  \state{Virginia}
  \country{USA}
  \postcode{23529}
  \\\url{https://orcid.org/0000-0002-2787-7166}
}

%\author{Martin Klein}
%\orcid{0000−0003−0130−2097}
%\affiliation{%
%  \institution{Los Alamos National Laboratory}
%  \streetaddress{P.O. Box 1663}
%  \city{Los Alamos}
%  \state{New Mexico}
%  \country{USA}
%  \postcode{87545}
%}
%\email{mklein@lanl.gov}

\author{Martin Klein}
% \affiliation{Los Alamos National Laboratory}
% \affiliation{Los Alamos, NM, USA}
% \affiliation{\city{Los Alamos}\state{New Mexico}\country{USA}}
% \affiliation{\url{https://orcid.org/0000-0003-0130-2097}}
\email{mklein@lanl.gov}
\affiliation{%
 \institution{Los Alamos National Laboratory}
 \streetaddress{P.O. Box 1663}
 \city{Los Alamos}
 \state{New Mexico}
 \country{USA}
 \postcode{87545}
 \\\url{https://orcid.org/0000-0003-0130-2097}
}

\author{Michael L. Nelson}
% \affiliation{Old Dominion University}
% \affiliation{Norfolk, VA, USA}
% \affiliation{\country{USA}}
% \affiliation{\city{Norfolk}\state{Virginia}\country{USA}
% \url{https://orcid.org/0000-0003-3749-8116}}
\email{mln@cs.odu.edu}
\orcid{0000−0003−3749−8116}
\affiliation{%
  \institution{Old Dominion University}
  \streetaddress{1 Old Dominion University}
  \city{Norfolk}
  \state{Virginia}
  \country{USA}
  \postcode{23529}
  \\\url{https://orcid.org/0000-0003-3749-8116}
}
% \email{mln@cs.odu.edu}

\renewcommand{\shortauthors}{Jones, et al.}

%%
%% The abstract is a short summary of the work to be presented in the
%% article.
\begin{abstract}
%To allow previewing a page before clicking on it, social media platforms have developed social cards: visualizations consisting of, at a minimum, a title, text summary, striking image, and domain name. Each of these features presents a different piece of information about the underlying resource. While platforms routinely generate social cards for live web resources, they are often not correctly generated for \emph{archived} web pages, including pages that lack or predate standards for specifying striking images.  In this work, we show how 40.44\% of archived news articles from our NEWSROOM dataset sample fail to supply striking images for social cards. From our PubMed Central dataset sample, 73.98\% of scholarly articles do present striking images, but they are repeating image patterns or logos of publishers and journals, images that do not summarize their article content. Due to these deficiencies, any service generating a social card for these types of resources will need to automatically generate striking images to summarize the underlying content.  The COVID-19 emergency has made it clear that scholarly articles are at an aesthetic disadvantage in social media platforms when compared to their often more flashy disinformation rivals.  Using a 37,552 news articles from NEWSROOM and 227,265 articles from the scholarly journal \emph{PLOS ONE}, we demonstrate that we can predict the striking image that best describes a document with a $Precision@1$ of 0.8825 for news articles and 0.7786 for scholarly publications.
To allow previewing a web page, social media platforms have developed social cards: visualizations consisting of vital information about the underlying resource. At a minimum, social cards often include features such as the web resource's title, text summary, striking image, and domain name. News and scholarly articles on the web are frequently subject to social card creation when being shared on social media. However, we noticed that not all web resources offer sufficient metadata elements to enable appealing social cards. For example, the COVID-19 emergency has made it clear that scholarly articles, in particular, are at an aesthetic disadvantage in social media platforms when compared to their often more flashy disinformation rivals. Also, social cards are often not generated correctly for \emph{archived} web resources, including pages that lack or predate standards for specifying striking images. With these observations, we are motivated to quantify the levels of inclusion of required metadata in web resources, its evolution over time for archived resources, and create and evaluate an algorithm to automatically select a striking image for social cards. We find that more than $40\%$ of archived news articles sampled from the NEWSROOM dataset and $22\%$ of scholarly articles sampled from the PubMed Central dataset fail to supply striking images. We demonstrate that we can automatically predict the striking image with a $Precision@1$ of $0.83$ for news articles from NEWSROOM and $0.78$ for scholarly articles from the open access journal \emph{PLOS ONE}.
\end{abstract}


%%
%% Keywords. The author(s) should pick words that accurately describe
%% the work being presented. Separate the keywords with commas.

%% A "teaser" image appears between the author and affiliation
%% information and the body of the document, and typically spans the
%% page.
% \begin{teaserfigure}
%   \includegraphics[width=\textwidth]{sampleteaser}
%   \caption{Seattle Mariners at Spring Training, 2010.}
%   \Description{Enjoying the baseball game from the third-base
%   seats. Ichiro Suzuki preparing to bat.}
%   \label{fig:teaser}
% \end{teaserfigure}

\maketitle

\begin{figure}
  \includegraphics[width=0.5\textwidth]{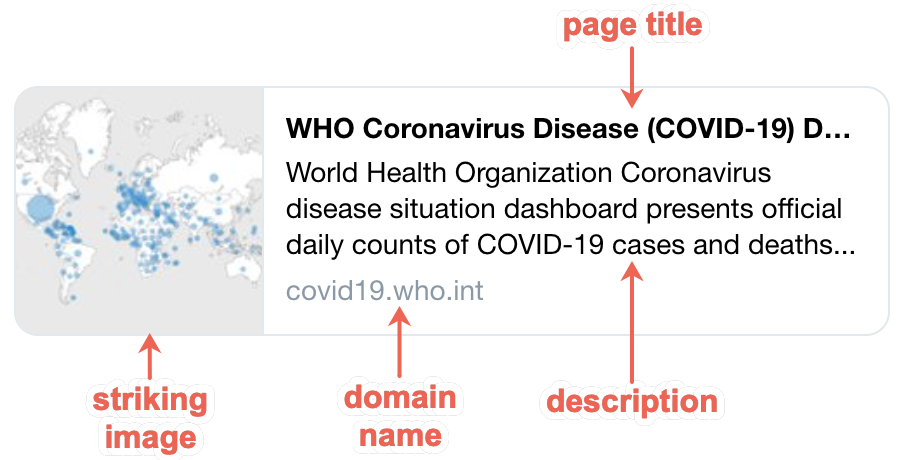}
  \caption{An annotated social card from Twitter for the URL \url{https://covid19.who.int/} identifying the social card units used in this paper.}
  \label{fig:diagram-of-social-card}
\end{figure}

\begin{figure}
  \includegraphics[width=0.5\textwidth]{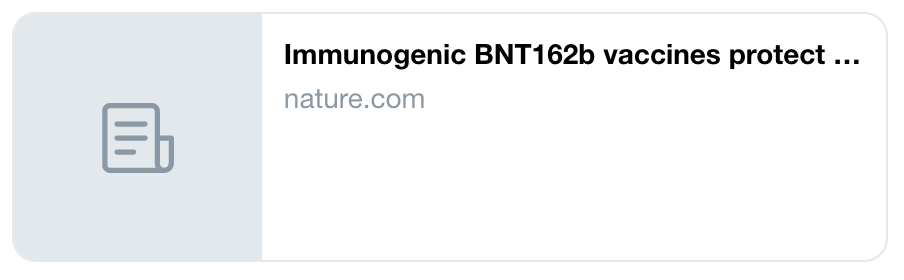}
  \caption{A social card from Twitter for the \emph{Nature} article URL \url{https://www.nature.com/articles/s41586-021-03275-y}, showing a lack of description or image.}
  \label{fig:academic-paper-bad-card}
\end{figure}

\begin{figure}
  \includegraphics[width=0.5\textwidth]{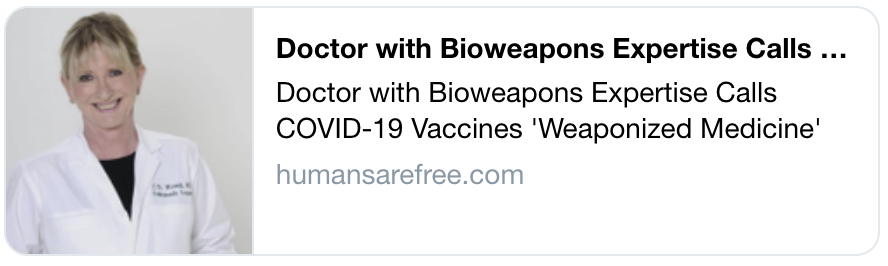}
  \caption{A social card from Twitter for the disinformation URL \url{https://humansarefree.com/2021/02/doctor-with-bioweapons-expertise-calls-covid-19-vaccines-weaponized-medicine.html}, showing all social card units filled.}
  \label{fig:fake-news-good-card}
\end{figure}

\section{Introduction}

Understanding the content behind a URL is important for web discourse. Though users might infer meaning from URLs, summaries are more effective.  To provide these summaries, social media platforms provide \textbf{social cards}, as shown in Figure \ref{fig:diagram-of-social-card}. Social cards provide different pieces of information to summarize the underlying content. These \textbf{social card units} typically consist of a title, a small text summary, a striking image, and a domain name that, in aggregate, summarize the page behind a URL. Social cards are similar to snippets in search engine result pages (SERPs) but have a slightly different purpose. Where search engine result snippets typically are dynamically contextualized based on the query and answer the question of ``Will this meet my information need?'', social cards are static and address the question ``What does the underlying page contain?'' Both allow for previewing a page and allow pages to ``compete for clicks'' relative to rivals that might be present elsewhere on the page. Textual information is important, but a good striking image can also help readers better understand the underlying document \cite{jones_social_2019}. As shown in the Twitter card from Figure \ref{fig:diagram-of-social-card}, the title and description give the reader some idea of the information found in the underlying page, but the striking image provides the additional insight of a current map of COVID-19 infection locations, helping the user infer that not only is a visualization available at this resource, but that it is also updated to contain current information.

In 2010, Facebook and Twitter established HTML metadata standards so page authors could supply their own values for social card units. However, we have observed that not all web pages offer sufficient information to generate helpful cards. Compare the card in Figure \ref{fig:academic-paper-bad-card} showing no image or descriptive metadata with the one in Figure \ref{fig:fake-news-good-card} that contains content for all social card units. The latter is more alluring and provides disinformation, whereas the former is a \emph{Nature} article providing peer-reviewed research, illustrating the phenomena detailed in ``The Truth Is Paywalled But The Lies Are Free'' \cite{robinson_truth_2020}. To facilitate the spread of accurate information, social cards can help by making such content more accessible. In the future, scholars will review the COVID-19 pandemic via web archives and their captures of specific observations of web pages. In prior work \cite{jones_mementoembed_2020}, we analyzed more than 60 platforms and determined that none reliably generated social cards for archived web pages. The roughly 150 billion web pages captured by the Internet Archive before 2010 \cite{internet_archive_internet_2009} likely have none of the 2010 standardized metadata. In this work, we will show how 40\% of our sample of archived news articles and 22\% of our sample of scholarly articles fail to specify striking images. Even though it appears that 78\% of scholarly articles contain striking images, we note that 74\% of them reuse the same image among multiple articles and 52\% of journals reuse the same image for all of their articles --- often a publisher or journal logo that does not summarize the article content. The lack of metadata and use of unsuitable images requires social card creation tools to automatically generate descriptions and select striking images for these documents. This fact along with the observed lack of platform support for creating cards for archived pages, the large number of pages that predate these metadata standards, and personal experience with poor support for some scholarly articles (e.g., Figure \ref{fig:academic-paper-bad-card}) inspired the research in this paper.

% news articles and scholarly publications are often sources for social cards on social media

Some of the endeavors benefiting from this research are social media storytelling, carousels for content management systems, and news aggregation platforms. To quantify this problem effectively, we analyzed news articles and scholarly publications --- resources that have undergone editorial review and, presumably, received some care in their publication. News articles and scholarly publications are also frequent sources for social cards on social media. Thus, in the following research questions, we contrast results between news articles (stored in the Internet Archive) with open access journal articles:

\textbf{Research Question \#1 (RQ1)} - What are the distributions of HTML metadata elements (general and social card elements) in news articles (over time) and scholarly publications published on the web?

%To address this question for news articles, we sampled 277,724 articles from the NEWSROOM \cite{grusky_newsroom_2018} dataset representing news articles captured in web archives between 1998 and 2016. For scholarly publications, we sampled 100 articles each from 1109 journals in the PubMed Central (PMC) open access commercial use dataset \cite{national_library_of_medicine_open_2019} for a total of 110,900 scholarly articles. 

\textbf{Research Question \#2 (RQ2)} - What approaches and image features are best suited to automatically select striking images from news articles and scholarly publications, and do the approaches differ for both resource types?

% Some of the endeavors benefiting from this research are social media storytelling, carousels for content management systems, and news aggregation platforms.

\section{Background}
\label{sec:background}

As shown in Figure \ref{fig:diagram-of-social-card}, full social cards consist of a page title, striking image, domain name, and description. To allow authors to supply their own values for these units, both Facebook and Twitter have developed the respective Open Graph Protocol (OGP) \cite{facebook_open_2021} and Twitter card \cite{twitter_cards_2021} standards. To supply a value for a social card unit, a web page author inserts the key-value pairs into the attributes of an HTML \texttt{META} element (e.g., {\small \texttt{<META} \texttt{property=}\texttt{"og:image"} \texttt{content=}\texttt{"http://example/image.png">}}). Table \ref{tab:social_card_units} lists the standard fields to be used with each social card unit. Most fields are optional, and cards will omit missing data, as seen in Figure \ref{fig:academic-paper-bad-card}. From our experience, Twitter requires that \texttt{twitter:card} and \texttt{twitter:title} or \texttt{og:title} exist or it will not create a card. Facebook is more forgiving, generating a card consisting only of the content of a page's HTML \texttt{title} element and domain name if no metadata fields are specified.  With the exception of \texttt{twitter:card}, if any other \texttt{twitter:*} fields are missing, Twitter will use the values of a corresponding \texttt{og:*} field if present \cite{twitter_getting_2021}. Deprecated or undocumented equivalents, such as \texttt{twitter:image:url}, may also be interpreted by the corresponding social media service. If present, Facebook favors the value of \texttt{og:title} over the content of the page's \texttt{title} element. The OGP specification states that Facebook should generate a card containing a description if \texttt{og:description} is present, but based on our experience, Facebook will provide a description in its card only if both \texttt{og:description} and \texttt{og:image} are present. This paper analyzes metadata per the specification.

If the author wishes to specify a striking image, they apply the appropriate field and image URL value to their \texttt{META} element. If a striking image URL is not specified, then a social card creation tool has a few options. It could display a default image in the card, as shown in Figure \ref{fig:academic-paper-bad-card}, providing no meaning for the end user and indicating that it found no striking image metadata field. Alternatively, it could render the web page in a browser and capture it as a screenshot, an artifact also known as a browser thumbnail. Finally, a tool can evaluate all of the images from the page and choose one, a technique we refer to in this paper as \textbf{striking image prediction}. For example, Blogger applies the simple approach of choosing the first image found within a post's content. Our recent work demonstrated \cite{jones_social_2019} that social cards probably perform better than screenshots for understanding; thus, we are analyzing approaches for the last option, striking image prediction. To achieve the best results, we analyze two different types of documents that are not well represented in other striking image prediction studies.

\begin{table}[t]
  \caption{Social card units and their associated cards standards field keys}
  \label{tab:social_card_units}
  \footnotesize
  \begin{tabular}{@{}lll@{}}
  \toprule
  \textbf{Social Card Unit}    & \textbf{OGP (Facebook)}   & \textbf{Twitter Card} \\ \midrule
  title                        & \texttt{og:title}       & \texttt{twitter:title}         \\
  description                  & \texttt{og:description} & \texttt{twitter:description}   \\
  striking image               & \texttt{og:image}       & \texttt{twitter:image}         \\
  identify the resource        & \texttt{og:url}         & N/A                   \\
  specify the type of resource & \texttt{og:type}        & \texttt{twitter:card}          \\ \bottomrule
  \end{tabular}%
\end{table}

Archived web pages, or \textbf{mementos}, contain a document's original HTML and images downloaded at some time in the past, recorded as the memento's \textbf{memento-datetime} \cite{van_de_sompel_rfc_2013}. This memento-datetime represents when the archive captured the memento and when the archive observed its properties, allowing us to examine web authors' behavior in the past. It is not necessarily the same as the publication date since archiving can occur well after publication. Sometimes the capture process fails to fully render a page, causing missing stylesheets, images, or JavaScript when an end user revisits the memento, a phenomenon called \textbf{memento-damage} \cite{brunelle_not_2015}. We refer to the URLs identifying these unchanging mementos as URI-Ms. Each URI-M identifies a capture of a specific version of a changing, live web resource known as the memento's corresponding \textbf{original resource}, identified by a URI-R \cite{van_de_sompel_rfc_2013}. Because of the problems with reliably generating cards for mementos \cite{jones_mementoembed_2020}, we created the social card creation tool MementoEmbed. We will update MementoEmbed with the results from this paper.

For RQ1, we use the NEWSROOM dataset \cite{grusky_newsroom_2018} developed by Grusky et al. for evaluating automatic text summarization algorithms against news articles. NEWSROOM contains 1.3 million URI-Ms of news articles for which there are textual summaries present in the article's HTML \texttt{META} elements. These textual summaries may come from the \texttt{*:description} fields shown in Table \ref{tab:social_card_units} or they may be specified in the standard HTML \texttt{META} element \texttt{description} field. The NEWSROOM dataset represents captures from 1998 through 2016 of news articles from 29 news outlets.

Also for RQ1, we examine scholarly publications as a contrast with our results for news articles. A digital object identifier (DOI) is a persistent identifier for locating a scholarly article regardless of website redesigns, corporate publisher acquisitions, and other phenomena that lead to broken links. Our work focuses on the HTML landing pages and HTML articles of open access scholarly publications. We acquired the PubMed Central (PMC) open access commercial use dataset \cite{national_library_of_medicine_open_2019} consisting of 1.7 million open access articles formatted as XML or plain text files. From these files, we extracted the DOIs, dereferenced them to download their HTML counterparts, and then analyzed the results. These articles are not mementos but are current versions of these scholarly publications. While DOI resolution on the web is not always reliable \cite{klein_tpdl_2020}, we used the same request methods and HTTP clients to obtain consistent results. 

For RQ2, we reuse a subset of the NEWSROOM dataset for news articles. As we mentioned in the Introduction and will show in Section \ref{sec:metadata_scholarly_pubs}, most of the articles in the PMC dataset do not provide good ground truth for striking images. Instead, we use all 227,265 articles from the open access journal \emph{PLOS ONE} found in the PMC dataset for evaluating striking image prediction. We chose to analyze the articles from \emph{PLOS ONE} because their submission guidelines \cite{plos_one_plos_2020} encourage each article's author(s) to choose a striking image from their article to represent it. \emph{PLOS ONE} also has other benefits, such as standardized URL patterns for detecting figures, tables, and equations within the document. This capability allowed us to produce a more intelligent image prediction approach that could discard images such as the PLOS logo, ORCID logo, and advertisements.

% We did not use \emph{PLOS ONE} for our metadata analysis because 98.78\% of \emph{PLOS ONE} articles in our sample contain striking images.

\section{Related Work}

Automatic image selection has been applied to the reduction of a large set of images to a small set for building photo albums \cite{ozkose_diverse_2019}, selecting representative pages from historical manuscripts \cite{cornia_automatic_2018}, choosing the best key frame to represent a video \cite{dirfaux_icim_2000,ren_wacv_2020}, generating collages \cite{tan_imagehive_2012} and general image collection summarization \cite{tschiatschek_learning_2014}. Individual image selection has been applied to detecting specific categories of images, such as spam \cite{liu_self-adaptable_2014}, advertisements \cite{changsheng_sensitivity_2010}, or landscapes \cite{hayashi_landscape_2007}, and coarsely identifying specific principal image subjects, such as \emph{vehicle} or \emph{pet} \cite{kalva_web_2007}. None of these solutions attempt to find the striking image that summarizes a single web page.

In 2004, Hu and Bragga \cite{hu_categorizing_2004} analyzed the front page of 25 randomly selected news sites and classified the images into seven categories. A \emph{story image} provides a striking image for a set of news articles covering a specific story. \emph{Preview images} provide striking images for specific articles. \emph{Commercial images} are advertisements. \emph{Host images} provide a photograph of an author. \emph{Heading images} are navigational elements consisting of stylized text. \emph{Icon logos} provide branding for the whole news source or a specific feature of the publication. \emph{Formatting images} perform the function of shaping or arranging a page, and examples include transparent spacing images or graphical horizontal rules, largely an artifact of the limited formatting abilities of the HTML of that era. The authors manually annotated 899 images across these 25 front pages. Their SVM classifier achieved an accuracy of 92.5\% when combining the discrete cosine transforms of each image with the values of its color bands and the surrounding text's properties.

In 2006, Maekawa, Hara, and Nishio \cite{maekawa_image_2006} analyzed forty websites and categorized 3,901 images into eleven categories. They then applied a custom classifier to the problem and achieved an accuracy rate of 83.1\% across categories. Maekawa's goal was to classify images so that mobile browsers could avoid the unnecessary download of images that would not display well on the smaller displays of mobile devices. Many of their categories are similar to Hu's. Their solution relies on easy-to-calculate features like width, height, byte size, number of colors, content type, and aspect ratio, but they also include the number of images on the page with similar features and textual information. While their overall accuracy is 83.1\%, they do poorly for certain categories, such as an $F_1$ score of 0.458 for identifying buttons and 0.694 for advertisements. 

Li, Shi, and Zhang \cite{li_improving_2008} ran an experiment in 2008 that is more similar to our work. They were interested in identifying striking images for search results.  Their \emph{dominant} image is similar to our concept of a striking image. For ground truth, they randomly sampled 3,000 documents from a dataset of pages from msn.com, mit.edu, and cnn.com. They then asked three participants to label each image as \emph{dominant} or \emph{non-dominant}. With this training data, they applied a custom classifier to predict the class of each image on a page. They used the features of pixel size, aspect ratio, sharpness, contrast, number of colors, categorization of photo or graphic with or without a human face, content-type, position on the page, size of image compared to the size of the page rendered in a browser, number of images larger than this image on a page, if the image came from an external site, and if the image repeats across the same web site. Their classifier calculates a relevance score for each image on the page based on the user's query with an accuracy of 0.85.

Koh and Kerne \cite{koh_deriving_2009} took a different approach. It starts by analyzing the HTML document's DOM. It finds the deepest nodes and works its way back up to discover the nodes most likely to contain content. From here, they choose the largest image with the smallest aspect ratio based on empirically determined thresholds. Based on human labeling, their algorithm achieves an accuracy of 0.898 and an $F_1$ of 0.921 across datasets of web pages consisting of 239 news pages and 254 research pages.

% First, these studies predate the rise of social media after 2009, and social media brings new influences to page authoring behavior.

Our work differs from that of Hu, Maekawa, Li, and Koh in several ways. Where Hu analyzed the front pages of news sources, we work with individual articles (i.e., deep links within the site). Unlike Hu and Maekawa, we are trying to find a single striking image to summarize the page rather than classifying images into multiple categories. We do not manually label images to develop our ground truth dataset; instead, we rely upon the actual image selected as part of the document's editorial review process as found in the \texttt{og:image} or \texttt{twitter:image} fields. Unlike Li's work, our method does not require a search query for selecting a striking image. We are inspired by many of the features chosen by all of this prior work. Unfortunately, some of these features are impossible to calculate reliably for our chosen documents. Mementos offer many challenges. For example, analyzing the images available on other pages from the same website may be impossible because the web archive did not capture other pages of the same website. Additionally, features that require rendering in a browser, like comparing page size to image size, may fail due to memento damage. Koh and Kerne are the only ones in this list to consider anything like scholarly publications. They did not process journal articles or conference proceedings but instead analyzed web pages from \emph{Scientific American}, \emph{IBM Research}, and \emph{Los Alamos National Laboratory}. Their solution will only work for HTML documents that contain images of varying sizes and aspect ratios. Our \emph{PLOS One} dataset contains many images whose size and aspect ratio are very similar, making it challenging to apply their solution.

\section{Distribution of Metadata Elements}

To address RQ1 we sampled news articles from the NEWSROOM dataset and scholarly articles from the PMC dataset.

\subsection{News Articles}

\begin{table}[t]
  \caption{NEWSROOM sample data reduction for metadata availability analysis}
  \label{tab:newsroom_sample_analysis}
    \footnotesize
  \begin{tabular}{@{}lrr@{}}
  \toprule
                          & \textbf{Count}  & \textbf{Running Total} \\ \midrule
  Initial Sample          & 310,163 & 310,163        \\
  Connection Failures     & 734    & 309,429        \\
  404 Not Found           & 7,570   & 301,859        \\
  503 Service Unavailable & 355    & 301,504        \\
  429 Too Many Requests   & 110    & 301,394        \\
  405 Method Not Allowed  & 54     & 301,340        \\
  403 Forbidden           & 2      & 301,338        \\
  400 Bad Request         & 1      & 301,337        \\
  Processing failures & 4,447 & 296,890 \\
  Redirects to dates after 2016 & 19,164 & 277,724 \\
  Remaining For Analysis  &        & 277,724        \\ \bottomrule
  \end{tabular}
\end{table}

We discovered that the 1.3 million article NEWSROOM dataset was unbalanced with respect to domain name and memento-datetime. For example, NEWSROOM has 186,095 mementos from \texttt{nytimes.com} and 1,429 mementos from \texttt{economist.com}. In terms of memento-datetime year, NEWSROOM contained 19 mementos from 1998 and 279,232 from 2016. Most of the mementos in the dataset come from 2016, making up 23\% of the dataset, and the percentage of mementos for each year decreases (e.g., 19\% for 2015, 13\% for 2014, 10\% for 2012). OGP came into existence in 2010 \cite{haugen_abstract_2010} and Twitter will use that standard. Because the number of mementos in the NEWSROOM dataset are more heavily biased toward years closer to 2016 and we wanted to contrast metadata usage before and after card standards were published, we added all 90,570 NEWSROOM articles with memento-datetimes from 2009 and before to our sample. For those after 2010, we created a bucket for each domain and memento-datetime year. We randomly chose URI-Ms until we filled each domain/year bucket to a size of 1,307 --- the median size of all domain/year divisions after 2010. Our sample size after this process was 310,163 mementos.

We downloaded this sample in June 2020. To lessen the chance of being rate limited by the Internet Archive, we divided the URI-Ms in the sample into seven subsets and spread them across different servers in Amsterdam, Frankfurt, London, New York, Northern Virginia, San Francisco, and Toronto.  We felt confident in this approach because the Internet Archive presents the content it recorded and does not alter it for different geographic locations. To address failed downloads caused by rate limiting, we repeated the downloads once in July and again in August 2020.  We discovered that the downloads with HTTP status codes of 400, 403, 404, and 405 were indeed captures of pages with those status codes. Some mementos redirected to mementos captured after 2016, similar to behavior that Ainsorth et al. \cite{ainsworth_only_2015} reported in a different set of mementos. We removed all mementos captured after 2016. After resolving these issues, as shown in Table \ref{tab:newsroom_sample_analysis}, our downloaded NEWSROOM sample consisted of 277,724 mementos.

% It is likely that changes at the Internet Archive \cite{ainsworth_only_2015} since NEWSROOM was created in 2016 resulted in slightly different results than Grusky et al. encountered. 

\begin{table}[t]
\caption{The number of mementos in the NEWSROOM sample capable of generating different social card units.}
\label{tab:newsroom_card_capable}
\footnotesize
\begin{tabular}{@{}lll@{}}
\toprule
                              & \multicolumn{2}{c}{\textbf{\# of Capable Mementos for Platform}} \\
                              & \multicolumn{2}{c}{\textbf{and \% of Dataset}} \\\cmidrule(l){2-3} 
\textbf{Social Card Units}    & \textbf{Twitter}                & \textbf{Facebook}              \\ \midrule
title only                    & 125,201 (45.08\%)               & 277,724 (100\%)*               \\
title only from metadata & 125,201 (45.08\%)               & 189,275 (68.15\%)              \\
title and description only         & 123,386 (44.43\%)               & 169,468 (61.02\%)$\dagger$              \\
title, description, image     & 121,823 (43.86\%)               & 165,426 (59.56\%)              \\ \bottomrule
\multicolumn{3}{l}{* if \texttt{og:title} is missing, Facebook will use the HTML's \texttt{title} element} \\
\multicolumn{3}{l}{$\dagger$ our testing shows that Facebook does not currently support} \\
\multicolumn{3}{l}{    this configuration, only displaying a card with a title} 
\end{tabular}
\end{table}

\begin{figure*}[t]
\includegraphics[width=0.9\textwidth]{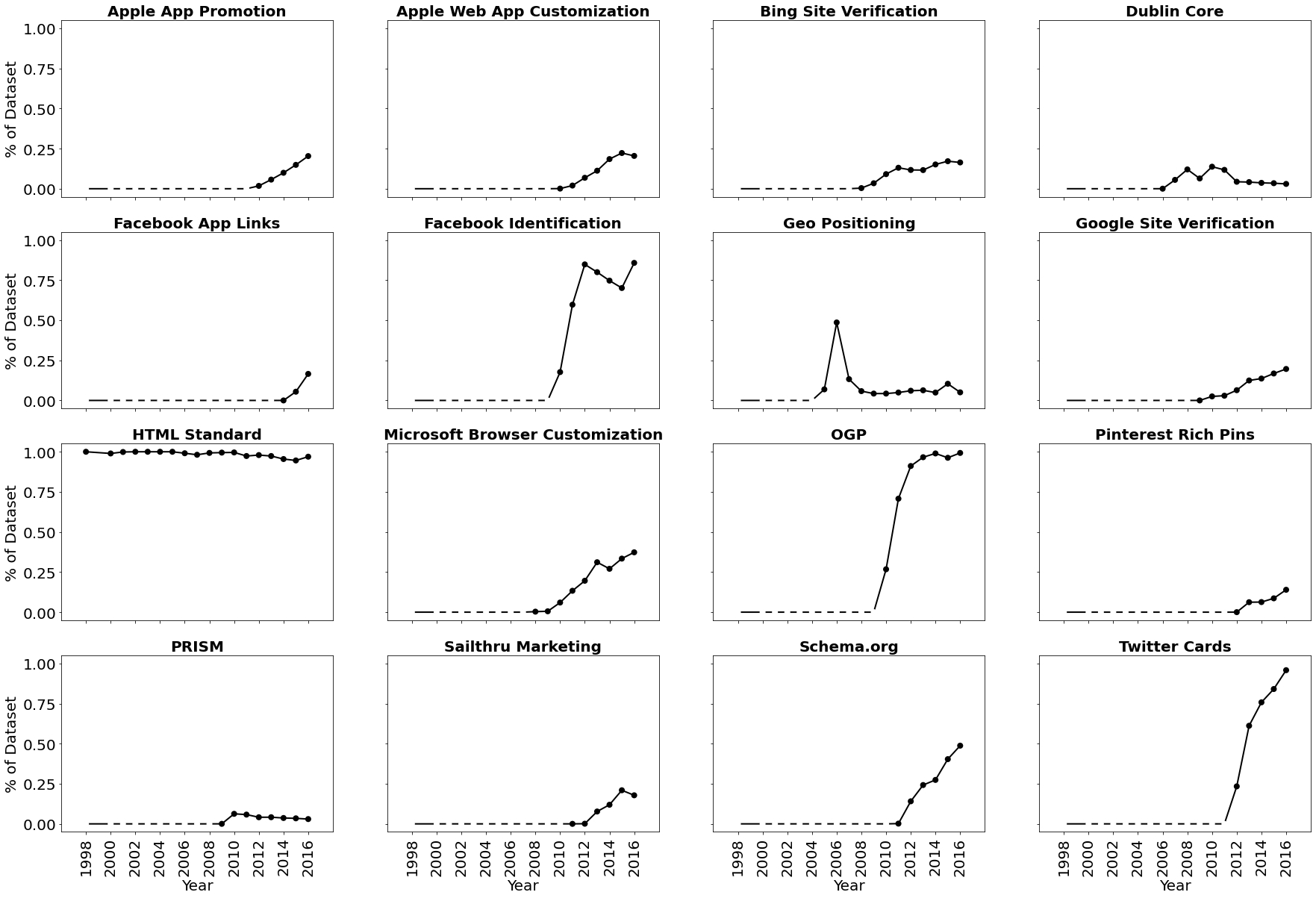}
\caption{By memento-datetime, the metadata categories of each memento from our NEWSROOM sample. Dashed lines indicate a y-axis value of 0.}
\label{fig:metadata-categories-over-time}
\end{figure*}

Table \ref{tab:newsroom_card_capable} lists the number of mementos in the NEWSROOM sample capable of creating different combinations of social card units. Because Facebook is more forgiving with missing fields, 59.56\% of articles can create a full card on Facebook, while only 43.86\% can do so with Twitter. We assigned each metadata field encountered to a category based on its corresponding standard or usage. We removed all instances where metadata fields were specific to a domain (e.g., only \texttt{nytimes.com} used the metadata field \texttt{byl}). Figure \ref{fig:metadata-categories-over-time} demonstrates how these categories changed over time. There is a focus on HTML standard metadata over time throughout our sample because all mementos in the NEWSROOM dataset contain at least a textual summary in some form, and before the OGP or Twitter standards, these articles used the HTML standard \texttt{description} field. We observed that news articles rapidly adopted social card metadata fields, starting with 13.13\% adoption of OGP in 2010 and reaching 93.05\% by 2016. After 2010, there is a rise in all types of metadata usage, focusing on search engines, mobile apps, browser customization, and social media, showing that news articles leveraged promotion of their content once standard metadata fields became available.

% schema.org started in 2011

\subsection{Scholarly Publications}
\label{sec:metadata_scholarly_pubs}

The PMC dataset of 1.7 million articles was organized by journal. To equalize the numbers per journal, we randomly sampled 100 DOIs from each journal's articles in the dataset. To avoid rate limiting, we downloaded these articles' landing pages from servers located in the English-speaking locations of London, San Francisco, New York, and Toronto. We ensured that we equally represented each journal at each location. Table \ref{tab:pmc_sample_data_reduction} shows how we were able to build a dataset of 110,900 articles in August 2020 for metadata analysis. This sample represents 1,109 journals from 209 publishers. These were not mementos but current web resources. Many of these articles are not stored in web archives, so we could not analyze metadata usage over time.

\begin{table}[t]
  \caption{PMC sample analysis data reduction}
  \label{tab:pmc_sample_data_reduction}
  \footnotesize
  \begin{tabular}{@{}lrr@{}}
  \toprule
                                                          & \textbf{Article} & \textbf{Running} \\
  \textbf{}                                               & \textbf{Count} & \textbf{Total} \\\midrule
  Initial Sample                                          & 128,604         & 128,604                 \\
  Failed Downloads                                        & 678            & 127,926                 \\
  404 Not Found                                           & 3,181           & 124,745                 \\
  401 Unauthorized                                        & 82             & 124,663                 \\
  503 Service Unavailable                                 & 6              & 124,657                 \\
  403 Forbidden                                           & 1              & 124,656                 \\
  520 (Cloudflare) & 1              & 124,655                 \\
  Journal Sample Contained Less Than 100 Articles                              & 13,755          & 110,900                 \\
  Remaining For Analysis                                  &                & 110,900                \\ \bottomrule
  \end{tabular}
\end{table}

% Please add the following required packages to your document preamble:
% \usepackage{booktabs}
\begin{table}[t]
\caption{The number of articles in the PMC sample capable of generating different social card units.}
\label{tab:pmc_card_capable}
\footnotesize
\begin{tabular}{@{}lll@{}}
\toprule
                              & \multicolumn{2}{c}{\textbf{\# of Articles Capable for Platform}} \\
                              & \multicolumn{2}{c}{\textbf{and \% of Dataset}} \\
                              \cmidrule(l){2-3} 
\textbf{Social Card Units}    & \textbf{Twitter}               & \textbf{Facebook}               \\ \midrule
title only                    & 44,845 (40.37\%)               & 110,900 (100\%)*                 \\
title only from metadata & 44,845 (40.37\%)               & 93,718 (84.51\%)        \\
title and description only         & 41,377 (37.31\%)               & 87,373 (78.79\%)$\dagger$                \\
title, description, image     & 40,750 (36.74\%)               & 86,346 (77.86\%)                \\ \bottomrule
\multicolumn{3}{l}{* if \texttt{og:title} is missing, Facebook will use the HTML's \texttt{title} element} \\
\multicolumn{3}{l}{$\dagger$ our testing shows that Facebook does not currently support} \\
\multicolumn{3}{l}{    this configuration, only displaying a card with a title}
\end{tabular}
% \vspace{-5mm}
\end{table}

Table \ref{tab:pmc_card_capable} summarizes our analysis of card capability within the PMC sample. Only 36.74\% of articles can create a full card, including striking image, on Twitter, while 77.86\% can do so on Facebook. Even though it appears that striking images for cards are well established for Facebook, 68,761/110,900 (73.98\%) articles reuse the same \texttt{og:image} value as another article. Most of these striking images were journal or publisher logos. In fact, 572/1109 (51.6\%) journals used the same URL for \texttt{og:image} with every article. Thus, few publishers use an image from the document's content and instead favor publisher or journal promotion. This contrasts with the NEWSROOM sample, where most striking images describe the article itself.

As we did with our NEWSROOM sample, we placed all metadata fields in the PMC sample into categories. Because these are current web resources, we could not show growth over time, so we ranked them instead. Facebook card metadata came in fourth place at 77.86\% of the dataset. Third place went to Dublin Core with 97,090/110,900 (87.55\%)  articles. Standard HTML fields like \texttt{keywords} and \texttt{referrer} came in second place with 102,704/110,900 (92.61\%) articles. In first place, 104,014/110,900 (93.79\%) articles applied metadata from Highwire Press, a metadata standard favored by Google Scholar \cite{acharya_highwire_2020}.

\section{Striking Image Prediction}

The numbers of mementos and scholarly articles lacking any meaningful striking image led to RQ2. Our overall goal was to find the approach that, given a set of images, will select the image closest to what a human selected, as found in the metadata, for the same document, regardless of whether or not that image also exists outside of the metadata. The striking images found in the metadata are our ground truth because the document's editorial process produced them.

% inspired us to examine how we might automatically select the best striking image from all candidate images in the document. 

% We evaluated how well a striking image prediction approach predicts the image found in metadata, whether or not that image also exists in the \texttt{BODY} of the HTML. 

% To determine how best to predict striking images we needed several things:
% \begin{enumerate}
% \item a dataset consisting of ground truth data
% \item an approach for selecting a striking image
% \item a method for evaluating the results of that approach run against the given dataset
% \end{enumerate}

Thus, any dataset applied to this endeavor requires documents where all images are available, and all documents must contain at least one striking image. This disqualifies documents that we cannot download or cannot parse with BeautifulSoup \cite{richardson_beautiful_2017} and documents with images that we cannot process with Pillow \cite{clark_pillow_2015} or ImageMagick \cite{the_imagemagick_development_team_imagemagick_2021}. We also disqualify documents with only one image because we have no prediction to make, and these easy wins may skew results. For each document, we consider image URLs found in the \texttt{META} elements from the \texttt{HEAD} as well as those provided by the \texttt{src} and \texttt{srcset} attributes of each \texttt{IMG} element found in the \texttt{BODY}.

% To find the images included in each document, our we must be able to download the documents and parse them with BeautifulSoup \cite{richardson_beautiful_2017}. For each document, we consider image URLs found in the \texttt{META} elements from the \texttt{HEAD} as well as those provided by the \texttt{src} and \texttt{srcset} attributes of each \texttt{IMG} element found in the \texttt{BODY}. For our experiments, we compute image features with the Pillow \cite{clark_pillow_2015} library. For later evaluation, ImageMagick \cite{the_imagemagick_development_team_imagemagick_2021} must be able to process all images. We eliminate documents that only contain a single image because there is nothing to predict. At least one image must be present in the HTML metadata as indicated by a value in the \texttt{og:image} or \texttt{twitter:image} fields. These images are our ground truth because the document's editorial process provided them to represent the document. Without at least one image specified by the metadata, we cannot evaluate if we correctly predicted the striking image for the given document.

For analyzing news articles, we started with our NEWSROOM sample of 310,163 mementos. As shown in Table \ref{tab:newsroom_reduction_for_image_study}, the issues we encountered left us with 37,522 articles to evaluate. The fact that we lose much of our dataset to download issues further emphasizes that many mementos can benefit from automatic striking image prediction because some of their images may be missing or corrupted.

\begin{table}[t]
  \caption{NEWSROOM sample data reduction for striking image prediction}
  \label{tab:newsroom_reduction_for_image_study}
  \footnotesize
  \begin{tabular}{@{}lrr@{}}
  \toprule
                          & \textbf{Article} & \textbf{Running} \\ 
                          & \textbf{Count} & \textbf{Total} \\ \midrule
  Initial Sample          & 310,163 & 310,163        \\
  Download Issues Detailed in Table \ref{tab:newsroom_sample_analysis} & 8,826 & 301,337 \\
  No Striking Image In Metadata & 188,612 & 112,725 \\
  Image Processing/Download Failures & 64,471 & 48,254 \\
  No Image In Body & 4,520 & 43,734 \\
  Only 1 image in body & 2,683 & 41,051 \\ 
  Cache failures  & 3,529 & 37,522 \\
  Remaining for Analysis &  & 37,522 \\ 
  \bottomrule
  \end{tabular}
\end{table}

As mentioned in Section \ref{sec:background} and demonstrated in Section \ref{sec:metadata_scholarly_pubs}, the PMC dataset was unsuitable for striking image prediction because the ground truth was poor or non-existent. We instead used a subset of the PMC dataset consisting of 227,265 \emph{PLOS ONE} articles because of their quality striking images. \emph{PLOS ONE}'s standardized URL patterns allowed us to produce a smarter image prediction approach that could discard images such as the PLOS logo or advertisements. We produced an HTML scraper that processes \emph{PLOS ONE} articles with these patterns in mind. We did not consider images found in supplemental sections or appendices. Table \ref{tab:plos-data-reduction} lists the issues we encountered when attempting to download and process these articles in July 2020, leaving us with 198,523 articles.

\begin{table}[t]
  \caption{PLOS dataset data reduction for striking image prediction}
  \label{tab:plos-data-reduction}
  \footnotesize
  \begin{tabular}{@{}lrr@{}}
  \toprule
                                                & \textbf{Article}  & \textbf{Running} \\
                                                 & \textbf{Count}  & \textbf{Total} \\ \midrule
  PLOS DOIs extracted from PMC dataset                                        & 227,265 & 227,265        \\
  Text   Corrections, not Articles                              & 5,840   & 221,425        \\
  Image or Metadata Corrections, not Articles                 & 207    & 221,218        \\
  Connection   Failure                              & 72     & 221,146        \\
  Parse   Failures                               & 15     & 221,131        \\
  No Image   In Metadata                         & 2,780   & 218,351        \\
  Pillow Image Processing Errors                  & 2,552   & 215,799        \\
  Pillow Conversion Failure                       & 1      & 215,798        \\
  Only 1 Image In Body                         & 1,129   & 214,669        \\
  Links do not match link text  & 8,373   & 206,296        \\
  Caption extraction failed     & 78     & 206,218        \\
  Unreferenced images            & 3,134   & 203,084        \\
  Striking image has no caption & 4,561   & 198,523        \\ 
  Remaining for analysis & & 198,523 \\
  \bottomrule
  \end{tabular}
\end{table}

Once we had our datasets, we applied different prediction approaches. Each prediction approach required one or more image features (e.g., byte size) to be successful. Some approaches choose the image with the highest or lowest value for the feature (e.g., largest or smallest byte size). For comparison with all approaches, we also ran 20 trials where we randomly chose each document's striking image.

\textbf{Base features}. Because social card creation tools must operate in real-time, we considered image features that could be calculated quickly. The Pillow library provides the following features for all image formats:
\begin{itemize}
  \item image byte size
  \item image width in pixels
  \item image height in pixels
  \item the number of columns in the image's histogram with a value of 0 (\textbf{negative space})
  \item image size in pixels
  \item the image's aspect ratio
  \item the number of colors in the image
\end{itemize}

In an attempt to achieve better results, we also supplied multiple features as input to various scikit-learn \cite{pedregosa_scikit-learn_2011} classifiers. When training classifiers, we placed images found in the HTML metadata into the class of \emph{present-in-metadata} and other images into the class of \emph{other}. When testing classifiers, we considered each document to be its own test case. We supplied the features of each image in a document to the classifier and asked it to provide the probability that the image comes from the class \emph{present-in-metadata}. From that set, we choose the image that has the highest probability of being in the class \emph{present-in-metadata} as the predicted striking image for that document. We chose this probability method so that each document contains at least one striking image prediction. If we had merely evaluated how well the classifier predicted an image's class (\emph{present-in-metadata} or \emph{other}), there would be documents for which the classifier found no striking image. With our method, even a low probability of belonging to \emph{present-in-metadata} still predicts a striking image because all of the document's images' probabilities are compared.

Social card creation tools have to contend with many different types of images. The values for some image features, like byte size, have no upper bound, making proper scaling challenging to estimate for the long term. Scaling can also remove precision from some measures, leading to poor results. Thus, we only considered the following classifiers because their scikit-learn implementations provide class probability scores and they do not require scaling:
\begin{itemize}
\item AdaBoost
\item Decision Tree
\item Gaussian Naive Bayes
\item Linear Discrimant Analysis
\item Logistic Regression
\item Random Forest
\end{itemize}

% % Instead, we want to know how well a given prediction approach chooses the correct striking image from all of the images found in a given document. 

We evaluate classifier operation with 10-fold cross-validation by document. Instead of evaluating how well each prediction approach predicts an image's class (\emph{present-in-metadata} or \emph{other}), we instead evaluate how well, given a set of images found in a document, the approach selects the ground truth striking image for that document. This leads us away from considering metrics like $F_1$ because they invite discussions about what recall means in this application. Instead, we consider other metrics commonly associated with information retrieval because we consider each document a query and the output of a prediction approach as a set of ranked results where the striking images are the relevant results. Thus, we apply $Precision@1$ ($P@1$) to determine the given approach's level of success for predicting the striking image from the metadata. If an approach produces a relevant result as its first result for a document, the $P@1$ score for that document and approach is 1, and 0 otherwise.  We take the mean of the $P@1$ scores for all documents, making $P@1$ a proxy for accuracy. We also supply Mean Reciprocal Rank ($MRR$) to evaluate how well an approach performed even if it failed to achieve $P@1=1.0$ for a document. For a given document, we determine the rank of the first relevant image ($rank_i$) as provided by the approach under test and compute the rank's reciprocal ($\frac{1}{rank_i}$). We then compute the mean of all of these reciprocal ranks across all documents for the same approach. For example, if 10 images exist in a document and the approach ranks a relevant image in fourth place, then the reciprocal rank is $\frac{1}{4} = 0.25$. A score of 1.0 for $P@1$ and $MRR$ is ideal.

We recognize that the same image may exist at different URLs or the striking image may be a cropped or resized form of the same image elsewhere in the document. For these issues, we apply a perceptive hash (pHash) distance to our evaluation as a proxy for human judgment of image similarity. If an approach selects image $A$, but the ground truth image is $B$, and if $pHashDistance(A, B) = 0$, then we consider $A$ to be just as relevant when computing $P@1$ and $MRR$. 

% We provide results at distances beyond 0 for those implementers that might accept higher distances of similarity.

Different pHash implementations exist. We evaluated ImageHash's pHash \cite{krawetz_looks_2011}, ImageMagick's pHash \cite{tang_perceptual_2012,weinhaus_tests_2014}, and Zauner's pHash \cite{zauner_implementation_2010}. We manually reviewed the intra- and inter-document similarity distances provided by each pHash implementation for the images found in 10 news and 10 scholarly articles. We concluded that ImageMagick's pHash provided the most intuitive similarity distances because it placed photographs at the same distance to each other, separate from logos and text. ImageMagick's pHash was more consistent with considering cropped or resized images to be similar to their original form. ImageMagick's pHash scores are not scaled and reached values as high as $6000$ during this evaluation. We noted that, while low scores intuitively indicated similar images, as scores reached higher values there is a greater discrepancy between the scores and human perception, thus we needed an upper bound for scaling that kept this discrepancy in mind. Our median score from this 20 article evaluation was 280.904. We scaled all distance scores such that scores above twice the median ($561.808$) became 1.0 and other scores are the result of dividing their value by $561.808$. ImageMagick has an issue processing certain JPEGs \cite{kozhuharov_6911_2020}, so we converted all images to PNGs before computing the pHash distance.

\subsection{Results of Predicting Striking Images For News Articles}

\begin{figure*}[t]

\begin{subfigure}[t]{3.3in}
  \includegraphics[width=3.2in]{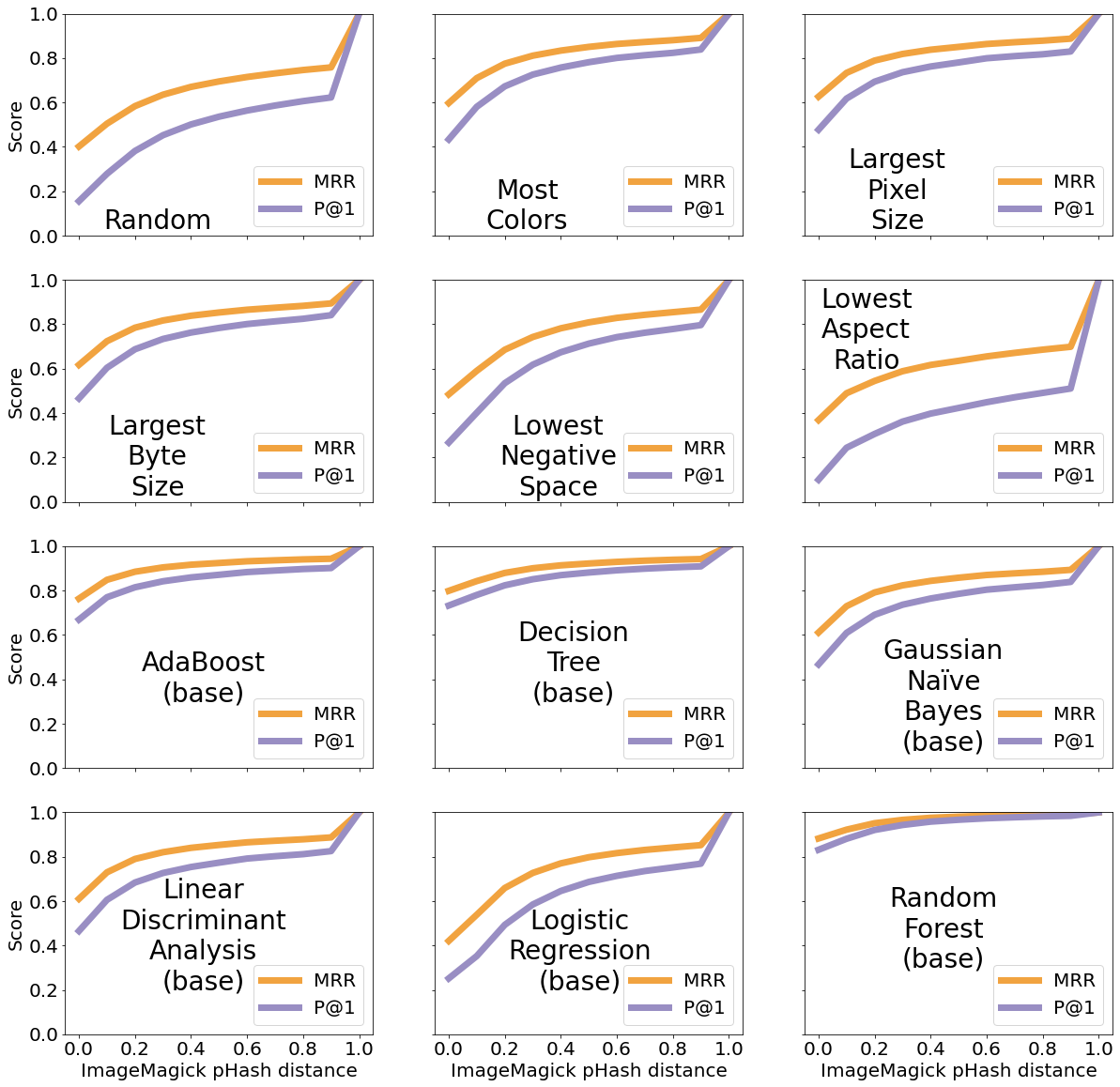}
  \caption{37,522 news articles from NEWSROOM}
  \label{fig:newsroom-prediction-results}
\end{subfigure}
\qquad
\begin{subfigure}[t]{3.3in}
  \includegraphics[width=3.2in]{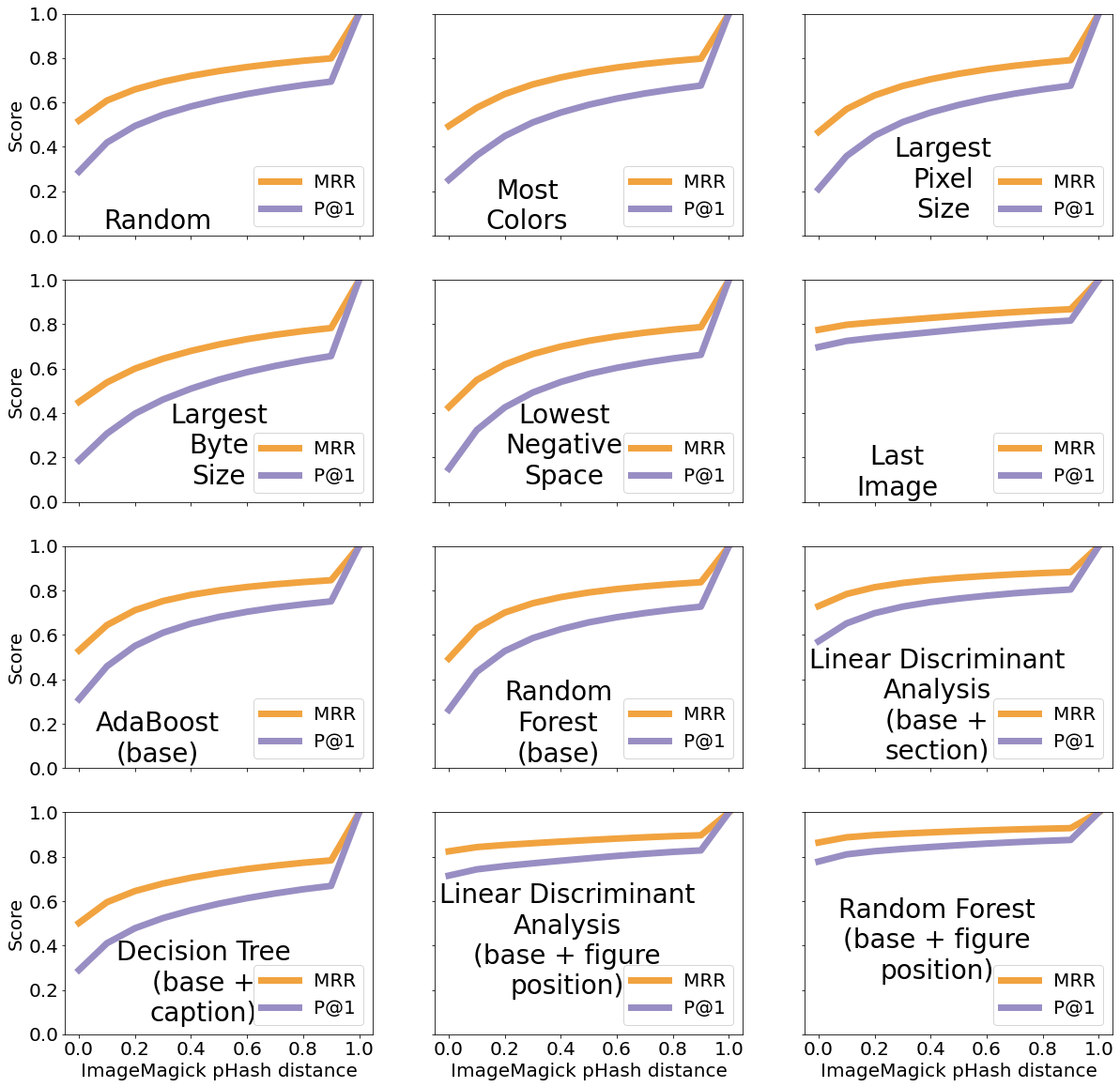}
  \caption{198,523 scholarly articles from \emph{PLOS ONE}}
  \label{fig:plos-image-prediction-results}
\end{subfigure}
\caption{These visualizations demonstrate the $MRR$ and $P@1$ results for different striking image prediction approaches as run against each dataset. The best approach achieves the highest $MRR$ and $P@1$ at the lowest pHash distance, making the ideal situation one where an approach's lines start higher into a graph's upper left corner. Items in parentheses are feature categories applied to the classifier, if applicable.}
\label{fig:all-image-prediction-results}
\end{figure*}

Figure \ref{fig:newsroom-prediction-results} demonstrates the performance of different prediction approaches with the NEWSROOM sample. Each line demonstrates the increasing $MRR$ or $P@1$ score, identified by its points' y-axis values, as produced by an evaluation that we performed at the corresponding pHash distance on the x-axis. When we evaluate at a pHash distance of 1.0, any image selected by the approach is equivalent to the ground truth; thus, $MRR$ and $P@1$ are 1. At a pHash distance of 0, only images that are perceptively equal to the ground truth image are relevant. The best approach achieves the highest $MRR$ and $P@1$ at the lowest pHash distance, resulting in lines that start in the upper left quadrant of each graph.

\begin{table}[t]
\caption{Correlation values for features in the NEWSROOM dataset to image present in metadata.}
\label{tab:feature_correlations}
\footnotesize
\begin{tabular}{llrr}
\toprule
 \textbf{feature category} & \textbf{feature}                    &   \textbf{Spearman's $\rho$} \\
\midrule
 base features & byte size                  &  0.2146  \\
              & width                      &  0.2191   \\
              & height                     &  0.2537   \\
              & negative space             & -0.1442   \\
              & size in pixels             &  0.2403   \\
              & aspect ratio              &  0.0225  \\
              & number of colors           &  0.1749  \\
\bottomrule
\end{tabular}
% \vspace{-15mm}
\end{table}

Randomly choosing images gives poor performance at a pHash distance of 0 with $MRR=0.4016$, $P@1=0.1551$. Training Random Forest with all base features results in the best performance at a distance of 0 with $MRR=0.8825$ and $P@1=0.8314$. To see if we could improve performance with fewer features, we applied Spearman's $\rho$ to our features, as shown in Table \ref{tab:feature_correlations}. We see that aspect ratio and negative space have the lowest correlation. We retrained Random Forest with these features removed and achieved $MRR=0.8782$, $P@1=0.8267$ at a pHash distance of 0. Removing additional features produced results with $P@1 < 0.8$.

%, meaning that we select the correct image $15.51\%$ of the time. 
% , producing the correct image 83.14\% of the time with the correct image, on average, at rank 1.133
%  which are close to the scores with these features included

\begin{table}[t]
\caption{Results for different classifier and feature combinations from our \emph{PLOS ONE} sample evaluated at a pHash distance of 0.}
\label{tab:plos_different_results}
\scriptsize
\begin{tabular}{@{}p{0.8cm}ccccccccc@{}}
\toprule
\textbf{}                    & \multicolumn{2}{c}{\textbf{base}}                                   & \multicolumn{2}{c}{\textbf{base + section}}                         & \multicolumn{2}{c}{\textbf{base + caption}}                         & \multicolumn{2}{c}{\textbf{base + figure position}}                  \\ \midrule
\textbf{}                    & \multicolumn{1}{c}{\textbf{MRR}} & \multicolumn{1}{c}{\textbf{P@1}} & \multicolumn{1}{c}{\textbf{MRR}} & \multicolumn{1}{c}{\textbf{P@1}} & \multicolumn{1}{c}{\textbf{MRR}} & \multicolumn{1}{c}{\textbf{P@1}} & \multicolumn{1}{c}{\textbf{MRR}} & \multicolumn{1}{c}{\textbf{P@1}} \\ \midrule
AdaBoost                     & 0.5300                           & 0.3102                           & 0.7202                           & 0.5533                           & 0.5086                           & 0.2824                           & 0.8316                           & 0.7096                           \\
Decision Tree                & 0.5204                           & 0.2911                           & 0.5513                           & 0.3564                           & 0.5019                           & 0.2895                           & 0.7602                           & 0.6653                           \\
Gaussian Naïve Bayes         & 0.4598                           & 0.2277                           & 0.6854                           & 0.5199                           & 0.4565                           & 0.2242                           & 0.7544                           & 0.6051                           \\
Linear Discriminant Analysis & 0.4763                           & 0.2469                           & 0.7298                           & 0.5714                           & 0.4739                           & 0.2462                           & 0.8245                           & 0.7151                           \\
Logistic Regression          & 0.4729                           & 0.2445                           & 0.6600                           & 0.4754                           & 0.4729                           & 0.2443                           & 0.5995                           & 0.4057                           \\
Random Forest                & 0.4932                           & 0.2621                           & 0.6698                           & 0.4770                           & 0.5118                           & 0.2824                           & \textbf{0.8643}                           & \textbf{0.7786}                           \\ \bottomrule
\multicolumn{9}{l}{For comparison:} \\
\multicolumn{9}{l}{random selection $MRR=0.5185$, $P@1=0.2883$} \\
\multicolumn{9}{l}{choosing last figure $MRR=0.7753$, $P@1=0.6975$} \\
\end{tabular}
% \vspace{-10.5mm}
\end{table}

\subsection{Results of Predicting Striking Images For Scholarly Publications}

Scholarly publications have additional features not available in web-based news publications. To improve our chances of discovering a meaningful striking image, we only included images that had captions and were directly cited in the paper. This decision excludes equations, which were never a striking image in the \emph{PLOS ONE} dataset. Equations as images is an artifact of HTML being incapable of reliably rendering mathematical notation. This left graphics and tables which are identified by \emph{PLOS ONE} with a \texttt{g} or \texttt{t} in their URLs, respectively. For simplicity, we refer to all captioned items, graphics or tables, as figures.

\textbf{Section features}. The first reference to a figure may indicate its importance to the paper. Our \emph{section index} feature is the number of the section where the image is first referenced. For example, in this paper, the first image is referenced in the Introduction, which has a section index of 1. The \emph{scaled section index} is a scaled version of this. \emph{Character position in section} and \emph{word position in section} are the positions of the earliest reference, by character or word, from any section where the figure is referenced in the paper.

\textbf{Caption features}. Performing text analysis on captions may indicate which figure is most important and thus a good candidate for the striking image. For each paper, we used NLTK to generate a list of all words, with stopwords removed, and then calculated their term frequencies within that paper. We computed a score for each caption by summing the term frequencies of that caption's words. We then ranked all captions in the paper by these scores, giving us the feature \emph{caption TF rank}; thus, a higher rank is a proxy for the caption's importance to the document. We scaled this across all captions in the paper for \emph{caption TF rank (scaled)}. Titles may contain insight into the meaning of a paper, so we computed the \emph{Jaccard distance between the words in the title and each caption}.

\textbf{Figure position features}. Finally, we processed each document to compute the order of figures. The \emph{figure position index} is the number of the figure as encountered in the HTML version of the document. We also computed a scaled version, \emph{figure position (scaled)}, based on the document's total number of figures.

Figure \ref{fig:plos-image-prediction-results} demonstrates the results of some of our striking image prediction approaches with the \emph{PLOS ONE} sample. As with Figure \ref{fig:newsroom-prediction-results}, the x-axis shows the pHash distance and the y-axis is the corresponding $MRR$ or $P@1$ score. Randomly selecting the image from a document performs at $MRR=0.5185$, $P@1=0.2883$ at a pHash distance of 0. Merely choosing the last figure's image achieves $MRR=0.7753$, $P@1=0.6975$ at a pHash distance of 0. Table \ref{tab:plos_different_results} provides detailed numbers for results with using different classifier-feature combinations. Our best scoring classifier-feature combination at a pHash distance of 0 was Random Forest trained with base features combined with figure position features at $MRR=0.8643$, $P@1=0.7786$. Applying the correlation information from Table \ref{tab:correlation-plos} to remove features produced results with $P@1 \le 0.7$. 

\begin{table}[t]
  \caption{Correlation Values for features in PLOS dataset to whether the image is present in metadata.}
  \label{tab:correlation-plos}
  \footnotesize
\begin{tabular}{lp{3.6cm}rrr}
  
  \toprule
  \textbf{feature category} & \textbf{feature}                    &   \textbf{Spearman's $\rho$} \\
  \midrule
   base features & byte size                                                        &  0.0336 \\
                & width                                                             & -0.0009  \\
                & height                                                            &  0.0762 \\
                & negative space                                                    & -0.0756 \\
                & size in pixels                                                    &  0.0761  \\
                & aspect ratio                                                & -0.0761    \\
                & number of colors                                                  &  0.0765   \\
   \midrule
   figure position & Figure position                                                &  0.2522 \\
   features & Figure position (scaled)                                              &  0.4727 \\
   \midrule
   section features & Section Index                         &  0.1161   \\
                  & Scaled section index                    &  0.1051 \\
                  & Character position in section &  0.1959  \\
                  & Word position in section  &  0.1950 \\
    \midrule
   caption features &    Caption TF rank                                                   & -0.0721 \\
   & Caption TF rank (scaled)                                          & -0.0086 \\
   & Jaccard of title and caption                                      & -0.0262 \\
   \midrule
   & Number of references                                              & -0.0373 \\
  \bottomrule
\end{tabular}
% \vspace{-4mm} % may need to be removed
\end{table}

\section{Future Work}

For those items in the NEWSROOM dataset that still exist on the web, it would be interesting to run our striking image prediction analysis on the current versions of those articles. Even though the articles are not current, their publishing platform has likely changed and may now produce card metadata.

Our results for \emph{PLOS ONE} may be specific to that journal, or they may be specific to the biology and medical fields serviced by PMC. Further striking image prediction analysis with different journals is needed to expand and generalize these results. The open access publisher \emph{Frontiers} supplies striking images for its HTML articles and may offer suitable comparison journals. Scholarly articles are more commonly published as PDFs, so we will re-evaluate this process with PDF extraction tools.

Additionally, another application of this research may be to suggest which images \emph{should} be included in the main sections of a paper instead of being relegated to its supplemental sections or appendices.

% We may also want to repeat the analysis of striking images in scholarly publications for DOIs from other datasets outside of of the PMC open access bulk dataset to ensure that the results hold. 

% Based on anectdotal evidence, e-commerce websites apply metadata fields to ensure that products and services display well on social media. Analyzing this metadata may be fruitful for understanding how these we can best summarize these resources and may provide additional insights into how to better market this content.

\section{Conclusion}

Social cards describe the content behind a URL, helping a user answer the question of ``What does the underlying page contain?'' A social card summarizes each web page through the units of page title, striking image, domain name, and description. While the title and domain name are often easily extracted, automatically generating the description and a striking image are more challenging. Social media platforms provide standards so that authors can insert their own values for these social card units into their HTML pages as metadata.

Per RQ1, our evaluation of the archived web pages (mementos) of 277,724 news articles from the NEWSROOM dataset revealed that card metadata fields are nonexistent for mementos captured before 2010. This translates into roughly 150 billion web pages at the Internet Archive for which social card creation tools will need to automatically generate descriptions and striking images. We found that news articles rapidly adopted social card metadata fields, with 13.13\% adoption in 2010 and reaching 93.05\% by 2016. We also evaluated 110,900 scholarly articles in the PMC open access dataset and discovered that while  77.86\% of scholarly articles specified a striking image, 73.98\% reuse the same image among multiple articles. In fact, 572/1109 (51.6\%) of journals used the same image in every article. Instead of selecting an image directly from the article to summarize it, the publications chose a colorful placeholder, or a journal or publisher's logo. This practice is not in line with non-scholarly publishers, and thus scholarly articles are at a disadvantage on social media.

This dearth of quality striking images inspired RQ2. By analyzing 37,522 articles from the NEWSROOM dataset and 198,523 articles from the journal \emph{PLOS ONE}, we have determined that the same automatic image selection approach cannot be applied to both types of documents. With NEWSROOM, we achieve $P@1=0.8314$, $MRR=0.8825$ at a pHash distance of 0 with Random Forest and the base features of image width, height, byte size, pixel size, negative space, aspect ratio, and color count. In our \emph{PLOS ONE} dataset experiment, applying these base features to Random Forest performed worse than randomly selecting an image. We achieved the best image selection performance of $P@1=0.7786$ and $MRR=0.8643$ at a pHash distance of 0 for \emph{PLOS ONE} with Random Forest by using the base features combined with the position of the figure on the page. We did not achieve better results with features like caption text or section references.

These results have implications for social media, news aggregation platforms, content management systems, and social media storytelling. We believe that better social cards will help platforms better summarize topics, equalize the probabilities that readers will select more informative content, and, most important of all, help readers understand the documents that others share with them.

\begin{acks}
The authors thank Jian Wu and Alexander Nwala for conversations that inspired parts of this research. We also thank Max Grusky for granting us access to the NEWSROOM dataset and PMC for making their Open Access Journal Dataset openly available.
\end{acks}

\bibliographystyle{ACM-Reference-Format}
\bibliography{references}

\end{document}